\documentclass{aa}
\usepackage{lscape}
\usepackage{rotating}
\input psfig.sty
\input epsf.sty

\begin{document}

\title{Gemini observations of Wolf-Rayet Stars in the Local Group Starburst Galaxy IC~10} 

\titlerunning{WR stars in IC~10}


\author{P.A. Crowther\inst{1} \and L. Drissen\inst{2} \and J.B. Abbott\inst{1} \and P. 
Royer\inst{3,4} \and S.J. Smartt\inst{5}}


     \institute{
Dept of Physics and Astronomy, University College London, 
Gower St, London WC1E 6BT, United Kingdom \and
D\'{e}partement de Physique, 
Universit\'{e} Laval, and Observatoire du
Mont M\'{e}gantic, Qu\'{e}bec QC G1K 7P4, Canada \and
Institut d'Astrophysique, Universit\'{e} de Li\`{e}ge, 5 ave de Cointe, 4000
Li\`{e}ge, Belgium \and
K.U.Leuven, Instituut voor Sterrenkunde, Celestijnenlaan 200B, B-3001
Leuven, Belgium
\and
Institute of Astronomy, University of Cambridge, Madingley Road, CB3 OHA, 
United Kingdom
}
\offprints{P.A. Crowther, \email{pac@star.ucl.ac.uk}}
\date{Received / Accepted}

\abstract{We present Gemini-N GMOS and CFHT MOS spectroscopy of Wolf-Rayet 
candidates in the Local Group dwarf galaxy IC~10  that were
previously identified by Massey et al. and Royer et al. From the
present spectroscopic survey, the WC/WN ratio for IC~10 remains unusually high, 
given its low metallicity, although none of the WC9 stars suspected from 
narrow-band imaging are confirmed. Our spectroscopy confirms 9 newly 
discovered Wolf-Rayet candidates from Royer et al., whilst spectral types of 14 
Wolf-Rayet stars previously
observed by Massey \& Armandroff are refined here.  In total, there are 26 
spectroscopically confirmed Wolf-Rayet stars in IC~10. All but one of the fourteen 
WC stars are WC4--6 stars, the exception being \# 10 from Massey et al.,
a broad-lined, apparently single WC7 star.  There are a total of eleven WN stars, which
are predominantly early WN3--4 stars, but include a rare WN10 star, 
\#8 from Royer et al. 
\#5 from Massey et al. is newly identified as a transition WN/C star.  Consequently, the 
WC/WN ratio  for IC10 is 14/11$\sim$1.3,  unusually high for a metal-poor galaxy.
Re-evaluating recent photometric data of Massey \& Holmes, 
we suggest that the true WC/WN ratio may not be as low as $\sim$0.3. Finally,
we present  ground-based finding charts for all confirmed WR stars, plus 
HST/WFPC2 charts for twelve cases.   
}
\maketitle

\section{Introduction}

Amongst Local Group galaxies, IC~10 is remarkable for its very 
high star formation rate, as inferred from an exceptional H\,{\sc ii} 
population (Hodge \& Lee 1990), large far-IR luminosity (Melisse 
\& Israel 1994), a non-thermal radio continuum  (Yang \& Skillman 1993) 
and a high surface density of massive stars (Massey \& Armandroff 1995).
These are direct 
evidence for a young, widespread burst of star formation. Indeed, IC~10
may be considered as a blue compact dwarf (Richer et al. 2001). This 
starburst phenomenon is also reflected in its large population of 
Wolf-Rayet stars, which is discussed here. 
IC~10 is about a factor of two smaller than the SMC, with
a comparable metallicity, log(O/H)=8.26 (Garnett 1990), 
and so represents another rare opportunity to 
study individual massive stars at low metallicity.

Despite its nearby distance of only 0.6--0.8Mpc (Kennicutt et al. 1998; Borissova et al. 
2000), the study of stellar populations of IC~10 is made difficult by its location
towards the Galactic plane, with a foreground reddening of $E(B-V)=0.77-0.85$  
(Sakai et al. 1999; Richer et al. 2001).
Massey et al. (1992) first highlighted the apparent massive 
star peculiarity of IC~10, suggesting the presence of 22 Wolf-Rayet 
stars from narrow-band  photometry. Massey \& Armandroff (1995) 
subsequently spectroscopically confirmed 15 stars, with a highly
unusual WC/WN ratio of 2, twenty times higher than the SMC 
(Massey \& Duffy 2001). This is 
puzzling, given that all other Local Group galaxies show a reasonably tight
correlation of decreasing WC/WN ratio with lower metallicity.
Very massive stars will evolve to the WC phase via the WN stage, whilst
less massive stars might not advance to the WC stage at reduced metallicity,
due to weaker winds throughout its evolution (Massey 2003). 

As Massey \& Johnson (1998) have argued, an exceptionally
 high WC/WN ratio for IC~10 might only be explained by (i)
an Initial Mass Function (IMF) 
skewed towards very massive stars, yet Hunter (2001) finds a 
normal
IMF for intermediate mass stars; (ii) an exceptional $\sim$5Myr old galaxy wide starburst 
which would produce the unusual WC/WN ratio (Schaerer \& Vacca 1998); (iii)
incompleteness amongst the WR population, especially amongst WN stars. It
is well known that WN stars are more difficult to detect from imaging surveys
since their emission lines are generally much weaker 
(e.g. Armandroff \& Massey 1985). 
%

Subsequent to the original Massey et al. (1992) survey,
Royer et al. (2001) used deep narrow-band imaging to suggest the 
presence of another 13 WR stars in IC~10. Amongst these candidates,
Royer et al. announced the detection of three WC9 stars. This was
totally unexpected, given that such late WC subtypes stars have only 
previously  been observed within the metal-rich regions of the Milky Way. 
All WC stars  in the low-metallicity Magellanic Clouds are of subtype WC4--5 
or WO (Breysacher et al. 1999), a sequence that IC~10 was expected to follow. 
Consequently, we have obtained follow-up
spectroscopy of most WR candidates in IC~10
using the Gemini-N facility multi-object spectrograph GMOS, with the remainder of
other candidates observed with the Canada France Hawaii Telescope (CFHT).
 The present 
paper discusses
our new observations in Sect.~\ref{obs}, presents a revised catalogue
of Wolf-Rayet stars in IC~10 in Sect.~\ref{catalog}.
Following the completion of our study, Massey \& Holmes (2002) have recently presented 
a new imaging survey of IC~10, which claimed to have resolved the hitherto unusual
WC/WN ratio via the discovery of many new (principally) WN stars. We discuss the
presents results in the light of this recent work in Sect~\ref{ratio}.

\section{Observations}\label{obs}

\subsection{Spectroscopy}

Spectroscopy of WR candidates in IC~10 were obtained 
with GMOS (Hook et al. 2002)
at the 8m Gemini-N on Mauna Kea during 
21 December 2001--16 January 
2002, via a joint Canadian-UK program, with principal
investigators LD and PAC. In order for the necessary slit 
masks to be prepared, a series of images  with 
GMOS was necessary. A series of four images, each with 100 sec exposures, 
were taken of two fields during poor seeing (1--1.5$''$) on 20 November 2001 
with the $g'$ Gunn filter.  Nevertheless, these were of sufficient quality for 
individual sources to be identified via comparison with
William Herschel Telescope narrow-band images obtained by Royer et al. (2001).

Approximately 15 WR candidates were simultaneously observed with GMOS
in each of the two overlapping fields -- centred at RA=00 20 28 and 
Dec=+59 17 40 (J2000, hereafter GMOS-UK) and RA=00 20 14 Dec=+59 18 25 
(GMOS-Can). 
In all cases, a  1.0 arcsec slit and B600 grating (centred at 560nm) were 
used, permitting a 2 pixel spectral resolution of 5\AA.
Total integration times of 5$\times$3600 s and 4$\times$3600 s were
obtained for targets in the GMOS-UK and GMOS-Can fields, respectively.
Two  Wolf-Rayet 
stars (Massey  \#5 and \#19) were common to both fields. Spare slits 
were used to observe sky regions, or H\,{\sc ii} regions from Hodge \& Lee (1990).
GMOS consists of three 2048 $\times$ 
4608 EEV CCDs (13.5$\mu$m pixels), such that the longer axis 
of the 6144$\times$4608 pixel focal plane is placed in the dispersion 
direction. 
(The detector array is close packed with 0.5mm gap between detector 
imaging areas.) 
Pipeline Gemini specific {\sc iraf} procedures were used to reduce the 
datasets. An internal CuAr arc lamp was used for wavelength calibration,
whilst a 60 s exposure of G191-B2B provided a flux calibration.

Since many of the WR candidates in IC~10 are closely spaced spatially, 
we were unable to observe all 28 candidates in our two overlapping GMOS
fields. Consequently, we observed a small subset of candidates with
MOS at the 3.6m CFHT on 18 October 2001,
using the 2048$\times$4500 EEV1 CCD. Three 1200s exposures were obtained
with the B600 grating at moderate airmass 1.5--1.7. Slits were cut, in
real time, to a slit width of 1.5$''$, providing a spectral resolution of
7\AA. Data were reduced and extracted
using standard {\sc iraf} procedures, except that wavelength calibration
was achieved via {\sc figaro}. No flux calibration was attempted for
the CFHT datasets.

\subsection{Photometry}

We have adopted continuum magnitudes for IC~10 WR candidates from 
Royer et al. (2001), using their dedicated filter set (Royer et al. 1998). 
Essentially, their c1 filter ($\lambda_c$=5055\AA, FWHM=51\AA) is close
to the usual WR narrow-band $v$ filter ($\lambda_c$=5160\AA, FWHM=130\AA)
defined by Smith (1968), 
whilst their c2 filter ($\lambda_c$=6047\AA, FWHM=35\AA) is comparable
to the WR $r$ filter ($\lambda_c$=6000\AA, FWHM=100\AA). In some cases, 
line-free continua were below the detection limit of the WHT. Unfortunately, our
Gemini GMOS and CFHT MOS imaging were 
not sufficiently deep for significant improvements upon this. 
In such cases, we were able to obtain estimates
of the $v$-band magnitude via our GMOS spectrophotometry. Slit losses
have not been taken into account, although agreement between narrow-band
photometry and spectrophotometry is excellent, with c1$-v$(GMOS) 
$\simeq$ 0.1 mag on average. This is expected given the zero-point scale
of the Royer et al. (1998) system.

We have also obtained V broad-band photometry of 10 confirmed and 2 candidate
WR stars from HST/WFPC2 images
obtained with the F555W filter (10 exposures of 1400s each; see Hunter 
2001).
The {\sc daophot} package within {\sc iraf} was used to obtain instrumental 
magnitudes, which were then calibrated following Dolphin (2000a, b).
We cross-checked our calibration by comparing Hunter's (2001) photometry 
of the dense clusters in the same images with our estimates, 
which agreed within 0.1 mag. In several cases, the  ground-based image
is resolved into multiple sources, which we shall refer to as
A (brighter source) and B (fainter source). A careful comparison
  of the F814W (I band), F555W (V) and F336W (U) WFPC2 images (see Hunter
  2001) shows that in all cases, star A is much bluer than star B. It
  is therefore very likely that star A is indeed the WR star in all
  cases. The only exception to this is WR24 from Massey \& Holmes
(2002) which will be discussed separately.
Again, for cases without HST V-band photometry, estimates were obtained 
from our GMOS spectrophotometry. For stars in common, 
$-$0.3 mag $\leq$ V(HST)$-$V(GMOS) $\leq$ 0.3 mag. 

We were also able to estimate interstellar
reddenings from c1$-$c2 (i.e. $v-r$), assuming an
intrinsic colour of $(v-r)_0=-$0.2 mag. Interstellar reddenings, $E_{b-v}$
are obtained via $E_{b-v} \sim 1.36 E_{v-r}$ using a standard 
extinction law. Note that ground-based 
photometry for [MAC92] 12
is heavily contaminated by nearby redder stars, 
as determined from UBV WFPC2 imaging, such that WHT photometry
suggests a reddening which is far too high. The same is true
for [MAC92] 14 such that we have set  $E_{b-v}$=1.0 mag for these in 
Table~\ref{wrcat}, which is typical of other nearby stars.

\begin{sidewaystable*}
\caption[]{Spectroscopically confirmed Wolf-Rayet stars in IC~10 from
catalogues of Massey et al. (1992), Royer et al. (2001) and Massey \& Holmes (2002).
Previous photometric and spectroscopic spectral types are taken from the 
above references, plus Massey \& Armandroff (1995). Broad-band V and narrow-band $v$ 
magnitudes are taken from archival WFPC2 imaging and Royer
et al. (their c1), respectively, except those in parenthesis, which are 
obtained from  GMOS spectrophotometry. See text for determination of 
absolute magnitudes. Line equivalent widths ($W_{\lambda}$) and FWHM refer 
to He\,{\sc ii} $\lambda$4686 for WN stars, and C\,{\sc iii}
$\lambda\lambda$4647--50/He\,{\sc ii} 
$\lambda$4686 plus 
C\,{\sc iv} $\lambda\lambda$5801--12 for WC stars.}
\label{wrcat}
\begin{tabular}{
l@{\hspace{3mm}}l@{\hspace{3mm}}l@{\hspace{4mm}}
l@{\hspace{4mm}}c@{\hspace{2mm}}
c@{\hspace{2mm}}c@{\hspace{2mm}}c@{\hspace{2mm}}
c@{\hspace{1.5mm}}c@{\hspace{2mm}}
c@{\hspace{1.5mm}}c@{\hspace{2mm}}
l@{\hspace{2mm}}l@{\hspace{2mm}}l@{\hspace{2mm}}l}
\noalign{\hrule\smallskip}
 Name  & RA & Dec & V & $v$ & $E_{\rm b-v}$ & $M_{v}$ & $\log 
W_{\lambda}$ & FWHM & $\log W_{\lambda}$ & FWHM &
\multicolumn{3}{c}{Spectral Type} & GMOS\\
\#       & \multicolumn{2}{c}{J2000} & mag & mag & mag & mag & 
\multicolumn{2}{c}{4650/4686} & \multicolumn{2}{c}{CIV 5801-12}
& Photometric &Old &New  & Field \\ 
\noalign{\smallskip\hrule\smallskip}
{}[MAC92] 1 & 00 19 56.97 & +59 17 08.0 &   &22.0& 1.02 &$-$6.0& 1.85 & 65 &
2.04 & 50 & WC &  & WC4-5+abs & Can,MOS\\ 
{}[MAC92] 2 & 00 19 59.65 & +59 16 55.3 &   &21.7& 1.17 &$-$7.0& 1.94 & 74 &
2.30 & 66 & WC? & WC & WC4 & Can,MOS\\ 
RSMV 6 &00 20 03.02 & +59 18 27.4  &(22.5)& 22.8 & 1.22 & $-$6.1 &2.25 & 
78 & 2.33 & 82  & WC &  & WC4       & Can \\
RSMV 5 &00 20 04.24 & +59 18 06.6  &(22.0)& 22.4 & 1.44 & $-$7.4 &2.09 & 
79 &
2.11 & 80 & WC  &  & WC4--5+abs & Can,MOS \\ 
{}[MAC92] 4 & 00 20 11.55 & +59 18 58.3 &(20.1)& 20.3 & 1.26 & $-$8.7 &1.70 
& 64 &
1.60 & 49 & WN? & WC  & WC4--5 & Can \\ 
{}[MAC92] 5 & 00 20 12.85 & +59 20 08.5 &(22.4)& 23.2& 1.14  & $-$5.3  &2.62 
& 53 &
3.10 & 79 & WN & WN  & WNE/C4   & UK,Can \\
 RSMV 13& 00 20 15.62 & +59 17 22.2 &(23.8)& (24.0) & (0.7:) & $-$2.8: & 
2.36 & 27 & 
& & WN  &  & WN5 & UK \\
RSMV 9 & 00 20 20.33 & +59 18 40.2 &(22.0)& 21.6 & 1.16 &$-$7.0 
&1.49 & 28 & 
& & WN7 &  & WNE+abs$^{c}$ & UK \\ 
 RSMV 8 & 00 20 20.56 & +59 18 37.8 &(20.8)& 20.8 &1.11  & $-$7.6 &0.85 &  
9 & 
& & WN8--9 &  & WN10   & Can \\ 
{}[MAC92] 7 & 00 20 21.87 & +59 17 41.5 &(19.4)& 19.6 & 1.01 & $-$8.4 &1.95 
& 84 &
1.91 & 82 & WN & WC & WC4--5+abs & Can \\
{}[MAC92] 9 & 00 20 22.60 & +59 18 47.3&23.1&(23.9)&(0.8:) & $-$3.2:& 2.38 & 
30 
& & & WN & WN  & WN3       & UK \\
 RSMV 11 & 00 20 22.68 & +59 17 53.9 &22.8&       &       &    &   &  &
& & WC & & WC4  & MOS \\
{} [MAC92] 10 & 00 20 23.36 & +59 17 42.6 &(20.9)& 22.2    & 0.98     & 
$-$5.7 
&3.06 & 68&
3.29 & 85 & WC & WC6--7 & WC7  & Can \\ 
{} RSMV 12 & 00 20 25.61 & +59 16 48.6 &22.7& (23.8) &(1.3:) & $-$5.4: & 2.31 
& 68 
& &  & WN &  & WNE & Can \\ 
{}[MAC92] 12 & 00 20 26.17 & +59 17 26.9 &21.9 & 22.0    & 1.0$^a$  & $-$6.0 
&1.86 & 58 &
1.84 & 49 & WN? & WC   & WC4+a  & UK \\
 RSMV 10 & 00 20 26.48 & +59 17 05.3 &22.9&       &       &          &  &  
&
 & & WC &  & WC4  & MOS \\
{} [MAC92] 13-A & 00 20 26.63 & +59 17 33.2 &20.8& 21.4   & 1.03 & $-$6.7 
&2.07 & 58 &
2.04 & 47 & WC & WC  & WC5--6 & Can \\
{} [MAC92] 13-B     & 00 20 26.62 & +59 17 33.4 &23.1& \\
{}[MAC92] 14-A & 00 20 26.87 & +59 17 20.2 &20.7& 21.4  & 1.0$^{a}$ & 
$-$6.6 &2.67 & 79 &
2.47 & 78 & WC & WC  & WC5 & UK \\
{} [MAC92] 14-B & 00 20 26.91 & +59 17 20.3 &22.4 & \\
{} [MAC92] 15 & 00 20 27.03 & +59 18 18.6 &   & 23.3 & 1.05  & $-$4.9   & &  
&
     &   & WN? & WC6--7 &  &         \\
{} [MAC92] 24-A & 00 20 27.67 & +59 17 37.7 & 18.8& & & &0.6:& 21 
& & &  & WN & WN+OB$^{c}$ &     \\
{} [MAC92] 24-B   & 00 20 27.82 & +59 17 37.5 & 21.3& & & &   &  
& & &    &                              &     \\
{} [MAC92] 24-C  &  & 00 20 27.75 & +59 17 36.3 & 22.0& & & &    & 
& & &    &  &     \\
{} RSMV 2-A  & 00 20 28.00 & +59 17 14.6 &21.4& 22.4   
& 1.24  & $-$6.5 &1.53 & 17 & & &  WN7--8 & & WN7--8 &  UK \\
{} RSMV 2-B &    & 00 20 28.04 & +59 17 14.9 &25.1&\\
{} [MAC92] 17-A  & 00 20 29.05 & +59 16 52.3 & 22.5 &  & & 
   &1.5:& 15 & & & WN & WN & WNE+OB &Can    \\
{} [MAC92]  17-B &     & 00 20 28.98 & +59 16 51.9 & 22.2 &      &      & 
   &    &    & & &    &     &       \\
{} [MAC92] 19-A & 00 20 31.01 & +59 19 04.5 &22.6& 23.3 & 1.32  & $-$6.0 
&2.07 & 25 
&
& & WN & WNE & WN4 & UK,Can\\
{} [MAC92]  19-B &    & 00 20 31.01 & +59 19 05.25 &24.0 &\\
{} [MAC92]  23 & 00 20 32.79 & +59 17 16.4 &   &       &      &          
&1.60 & 13 & 
&  & & WN7--8 &   & \\
{} [MAC92] 20 & 00 20 34.46 & +59 17 14.7 &22.0& (22.8)  & (0.7:) & 
$-$3.9:&2.95 & 62&
2.86 & 52 & WC & WC  & WC5 & UK \\
{} [MAC92]  21 & 00 20 41.64 & +59 16 25.3 &   & 23.4 & 0.76 & 
$-$3.6 &2.44 & 26 & 
 & & WN & WN & WN4 & UK \\
\noalign{\smallskip\hrule\smallskip}
\noalign{\smallskip\hrule\smallskip}
\end{tabular}
\newline ($a$): Set at $E_{b-v}$=1.0 due to severe contamination of ground based 
photometry by nearby red stars (see text);\newline
 ($b$) RSMV 9 does not appear to be spatially coincident with [MCA92] 6  -- see  Fig.~1
\newline 
($c$) Spectral type re-evaluated based on published spectroscopy (sect~3.2)
\end{sidewaystable*}

\begin{figure*}[htbp!]
\epsfysize=24cm \epsfbox[-15 10 525 800]{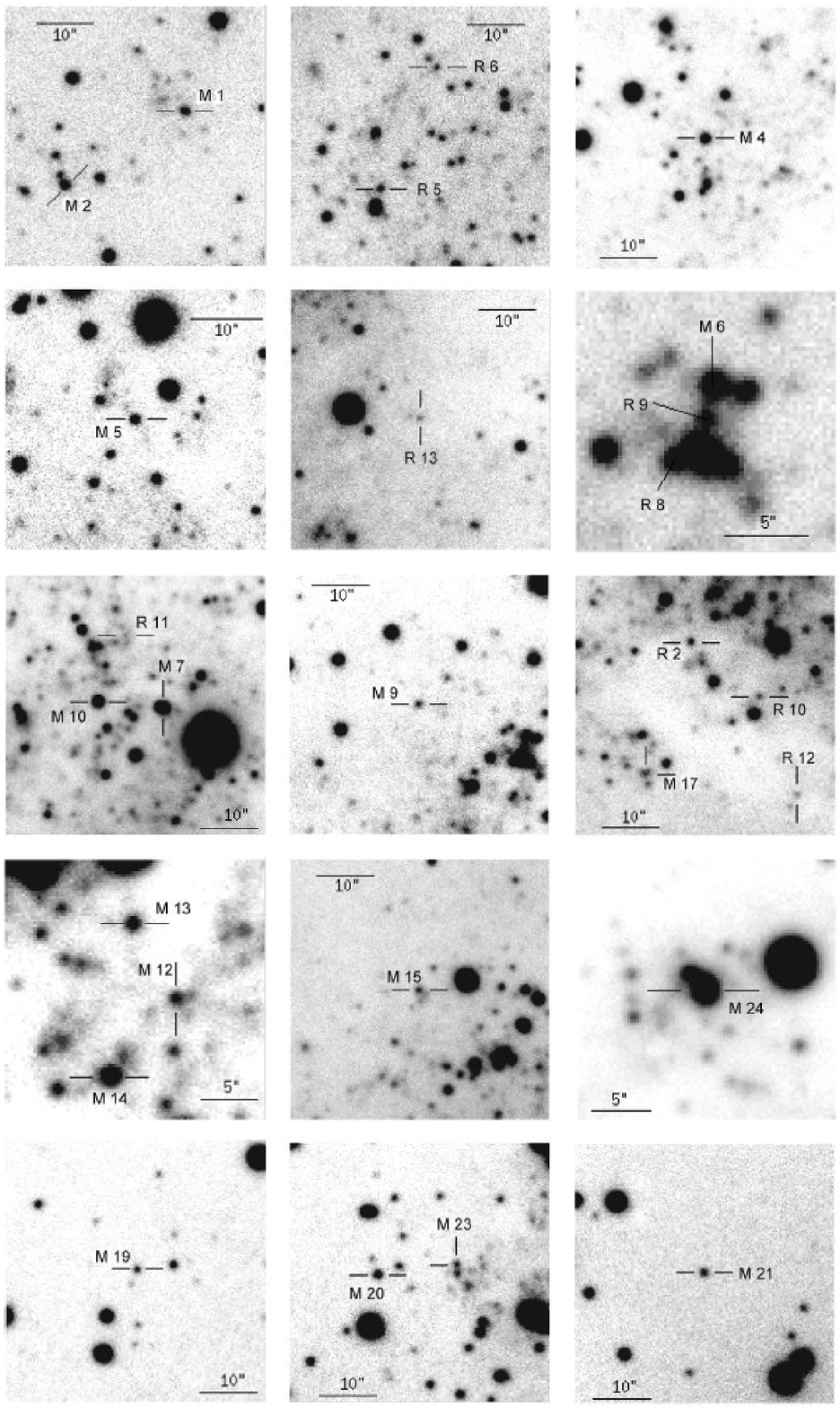}
\caption{Finding charts (typically 45$''$$\times$45$''$)
for WR stars in  IC~10 based on WHT He\,{\sc ii} narrow-band imaging.
Stars listed in the present work are indicated with  either M \# 
(MAC92) or R \# (RSMV).  [MAC92] 6 is spatially close (a few arcsec away) 
from the confirmed WR stars RSMV 9 and RSMV 8, but is not apparently coincident (despite
suggestions to the contrary by Massey \& Holmes 2002). North is up and east is to the left.}
\label{chart1}
\end{figure*}

\begin{figure*}[htbp!]
\epsfxsize=19cm \epsfbox[10 175 550 825]{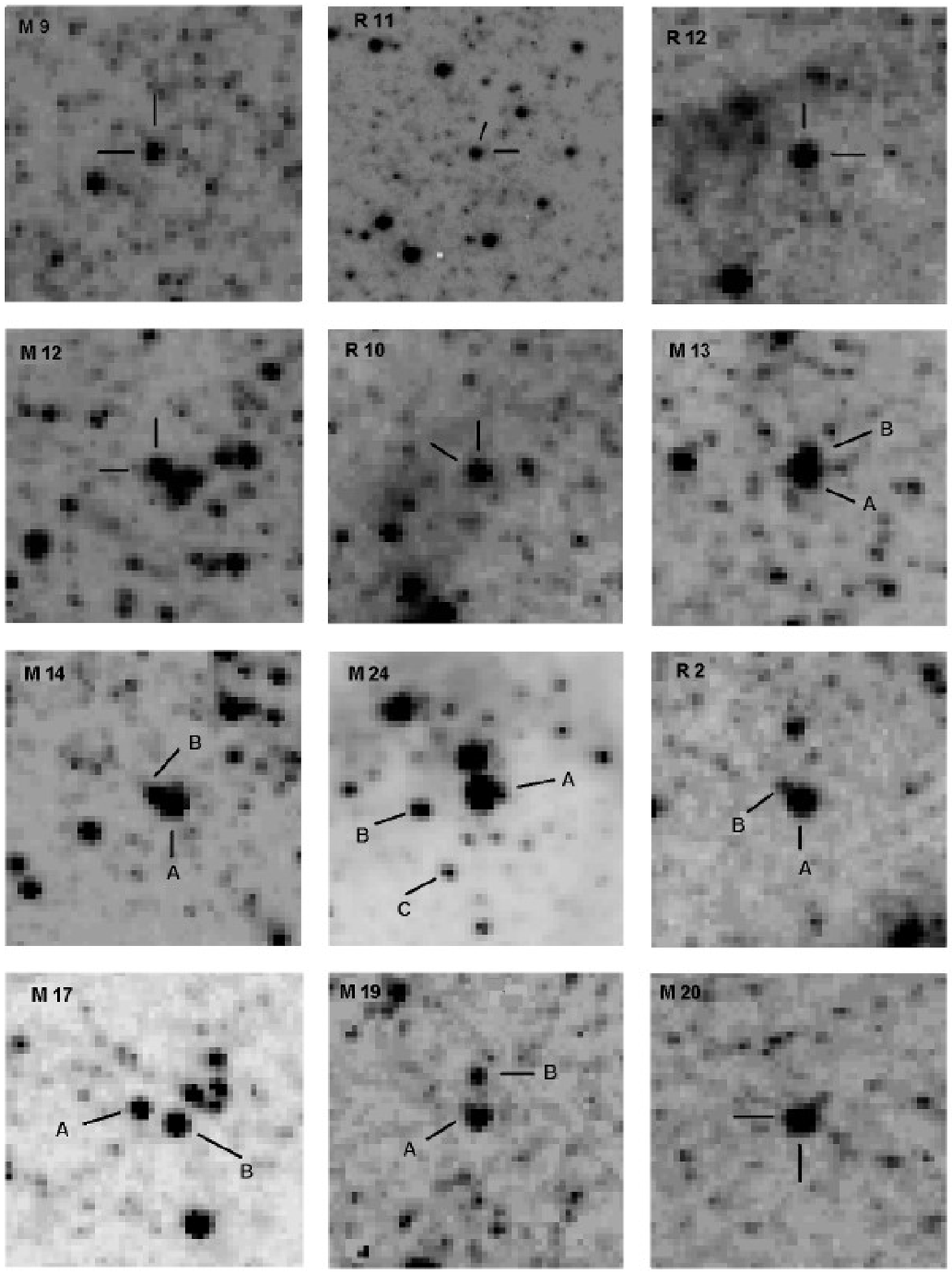}
\caption{Finding charts (5$''$$\times$5$''$) for selected 
WR stars in IC~10 based on HST WFPC2 F555W imaging.
As in the previous figure, catalogue names are from the
present work, and in some cases ground-based point sources are resolved 
into  multiple sources, -A and -B (or -C). In general, source A is the brighter and
bluer object, and so is presumably the WR star,
with the exception of [MAC92] 24,
for which the He\,{\sc ii} emission line star is uncertain. With this
exception,  both -A and -B  were included in our Gemini 
spectroscopy.  The pixel scale of RSMV 11 is different since it comes from the 
high resolution PC1 CCD (0.046$''$ instead of 0.1$''$). North is up and east is to the left.}
\label{chart2}
\end{figure*}

\subsection{Metallicity of IC~10}


As discussed above, we were able to observe several H\,{\sc ii} regions of 
IC~10 with GMOS, with a view to re-determining the metallicity of
this galaxy. Previously, Lequeux et al. (1979) 
obtained log(O/H)+12=8.17 for their H\,{\sc ii} region \#1 and 
8.45 for region \#2, whilst
Garnett (1990) re-determined 
log(O/H)+12=8.26, i.e. 0.25$Z_{\odot}$, for the
latter\footnote{We adopt log(O/H)+12=8.83 for the Solar abundance (Grevesse \&
Sauval 1998)}. This is intermediate between the oxygen content measured 
in
the Small and Large Magellanic Clouds. More recently,
Richer et al. (2001) determined a lower range of log(O/H)+12=7.84--8.23
for Hodge \& Lee (1990) H\,{\sc ii} regions 106b and 111b/c. 

From the 144 H\,{\sc ii} regions discussed by Hodge \& Lee, we have obtained 
GMOS spectroscopy for four - HL 16, 22, 45 and 128. In addition, we 
observed four candidate H\,{\sc ii} regions identified from narrow-band
imaging by Royer et al. (2001). We employ
[O\,{\sc iii}] $\lambda$4363/$\lambda$5007 for $T_e$ and
[S\,{\sc ii}] $\lambda$6717/$\lambda$6731 for $N_e$.
Unfortunately, far  blue spectroscopy necessary for the measurement of 
[O\,{\sc ii}] $\lambda$3727 is absent, such that merely lower
limits to the oxygen abundance may be obtained. Sadly, 
[O\,{\sc iii}] $\lambda$4363 lies below the detection 
threshold of most H\,{\sc ii} regions, so it ultimately 
proved difficult to determine
nebular conditions from our present set of GMOS observations.

Observations of HL 22 imply $E(B-V)$=0.99 mag from the observed 
$F$(H$\alpha$)/$F$(H$\beta$) ratio of 8.1, assuming a
standard intrinsic ratio of 2.85. Nebular diagnostics imply 
$T_{e}$=10,800K and $\log (N_e$/cm$^{-3}$)=2, such that the 
de-reddened $I$([O\,{\sc iii}] 5007)/$I$(H$\beta$) ratio of 3.31 reveals
log(O$^{2+}$/H$^{+}$)+12=7.97. 
From previous work, O/H$\sim$ 1.3 O$^{2+}$/H$^{+}$ 
(Lequeux et al. 1979), such that our observations of HL 22 suggest
log(O/H)+12$\sim$8.1, comparable to other recent determinations. 
 
\section{Properties of Wolf-Rayet stars}\label{catalog}

We have used GMOS/MOS spectroscopy of Wolf-Rayet candidates in IC~10
to verify previous results, in many cases from narrow-band photometry alone.
In most cases we have been able to confirm photometric spectral 
classifications, although there are some exceptions.

\subsection{New Catalogue}

The two previous narrow-band
imaging surveys in IC~10 by Massey et al. (1992, [MAC92]) and Royer et al. (2001, RSMV) 
led to the discovery of 22 and 13 candidates, respectively. 
Of the 22 Massey et al. candidates, four were confirmed in the original
paper, plus a further 11 were spectroscopically confirmed by Massey \& 
Armandroff (1995). New Gemini GMOS spectroscopy for 14/15 WR stars from Massey et al.
are presented here\footnote{[MAC92] 15 was not observed with either
Gemini or CFHT. Fortunately, the positive WR identification of this star is not in 
doubt, since  Massey \& Armandroff (1995) spectroscopically
demonstrate that this is a  WC6--7 star.}. WR classifications are supported in all 
cases, although 
[MAC92] 5 is  identified as a member of the rare WN/C intermediate subclass, due 
to exceptionally strong C\,{\sc iv} 5801--12\AA\ emission. This 
is evident in the spectrum presented by Massey \& Armandroff
(1995), and explains why narrow-band photometry of Royer et al. (2001) suggested a WCE identification.
We also obtained GMOS spectroscopy of two remaining candidate WR stars, \#3 and 16
from Massey et al, neither of which was confirmed as a Wolf-Rayet star.


The present study represents the first spectroscopic follow up to the 13 photometric
candidates of Royer et al. (2001). Ten candidates were observed with Gemini GMOS,
with the remainder (RSMV 1, 10 and 11) observed with CFHT MOS. In total, 9 stars were
confirmed as WR stars, with the exception of all three WC9 candidates (RSMV 1, 3 and 7),
plus the WC candidate RSMV~4.

Subsequent to the completion of our observing program, Massey
\& Holmes (2002) have spectroscopically confirmed two other emission line
stars in IC~10, although see Sect~3.2 with regard to the nature of \#24. 
In Table~\ref{wrcat}, we list 26 WR stars -- as is  usual ordered by increasing right 
ascension -- that are either confirmed here with our new spectroscopy, or literature 
datasets firmly establish their nature.
The present catalogue supersedes the earlier Massey et al. (1992) listing
(see also Table 12 of Massey \& Johnson 1998). Note that
`WR 4' from Massey et al. (1992), which is identified as IC~10 WR 4 by Massey \& Johnson (1998)
should be referred to as [MAC92] 4. Similarly, candidate WR star \#6 from Royer et al. (2001)
may be referred to as RSMV 6. Where high spatial resolution observations reveal multiplicity, 
we have denoted the components by A, B, etc. Use of this nomenclature is recommended to alleviate any possible confusion.
Indeed, should large numbers of bona-fide WR stars be identified in IC~10, as appears likely (Massey
\& Holmes 2002), a master catalogue may eventually need to be be created.
 



We present finding charts for all confirmed WR stars in IC~10 in Fig.~\ref{chart1}. 
These were obtained from our WHT He\,{\sc ii} $\lambda$4686 narrow-band 
imaging and supersede earlier charts from Massey et al. (1992). In 
Fig.~\ref{chart2}, we present HST WFPC2 images of 12 WR stars, revealing 
several that are resolved into multiple objects, A and B. 
In these instances, the Gemini GMOS slit contained both objects. 

\begin{figure}[hbtp]
\epsfxsize=8.8cm \epsfbox[25 570 280 685]{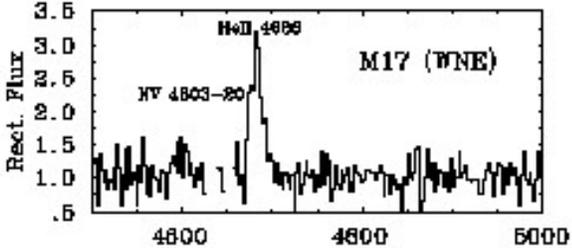}
\caption{GMOS spectroscopy of [MAC92] 17 from Massey et al. (1992),
in the vicinity of He\,{\sc ii} $\lambda$4686 indicating a WNE
spectral type.}
\label{ofwn}
\end{figure}

\begin{figure*}[htbp!]
\epsfxsize=17.5cm \epsfbox[25 285 520 720]{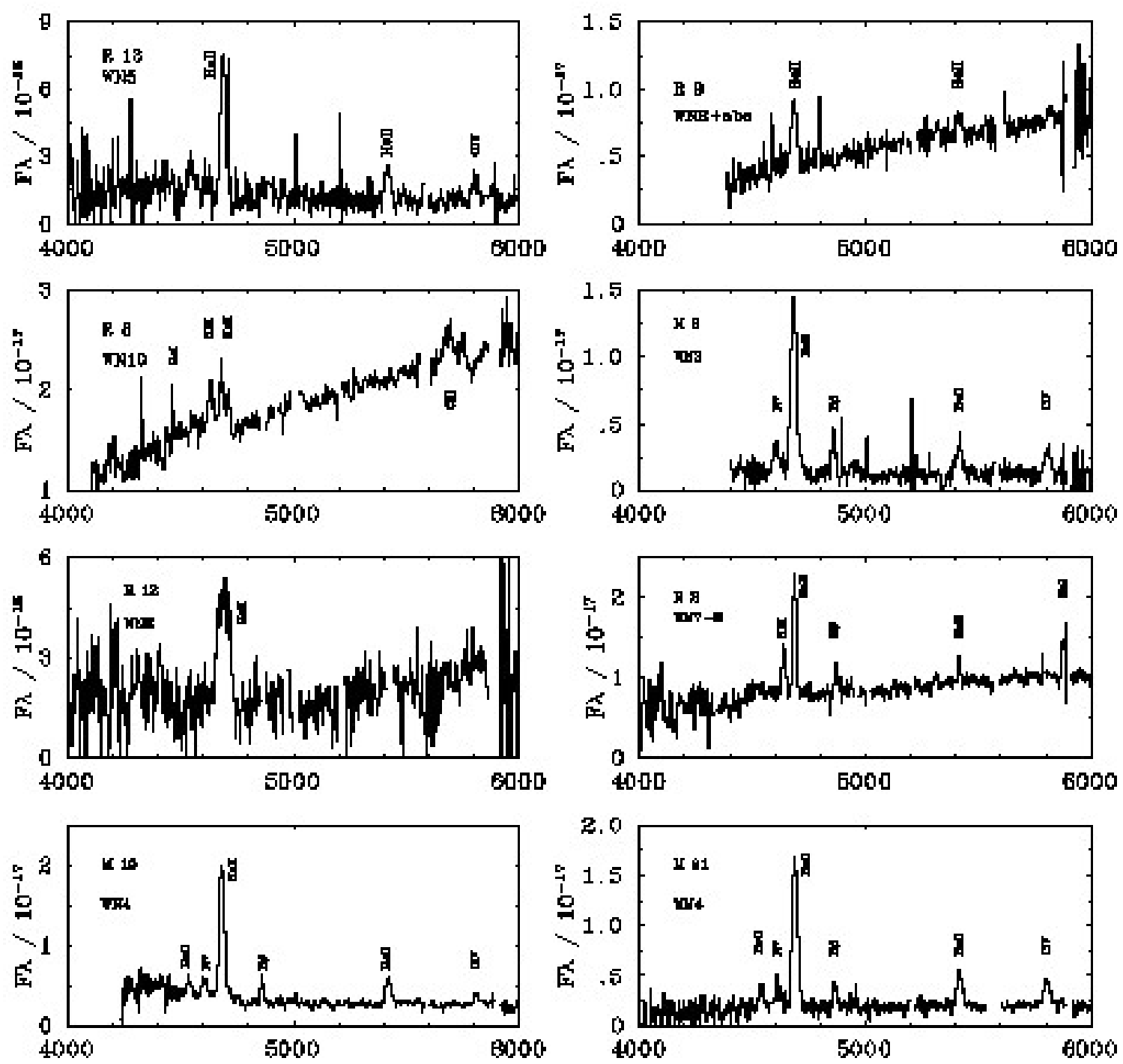}
\caption{Flux calibrated (erg\,cm$^{-2}$\,s$^{-1}$\,\AA$^{-1}$) optical 
spectroscopy of WN stars in IC~10 observed with Gemini GMOS.}
\label{WN}
\end{figure*}

\begin{figure*}[htbp!]
\epsfxsize=17.5cm \epsfbox[25 75 520 720]{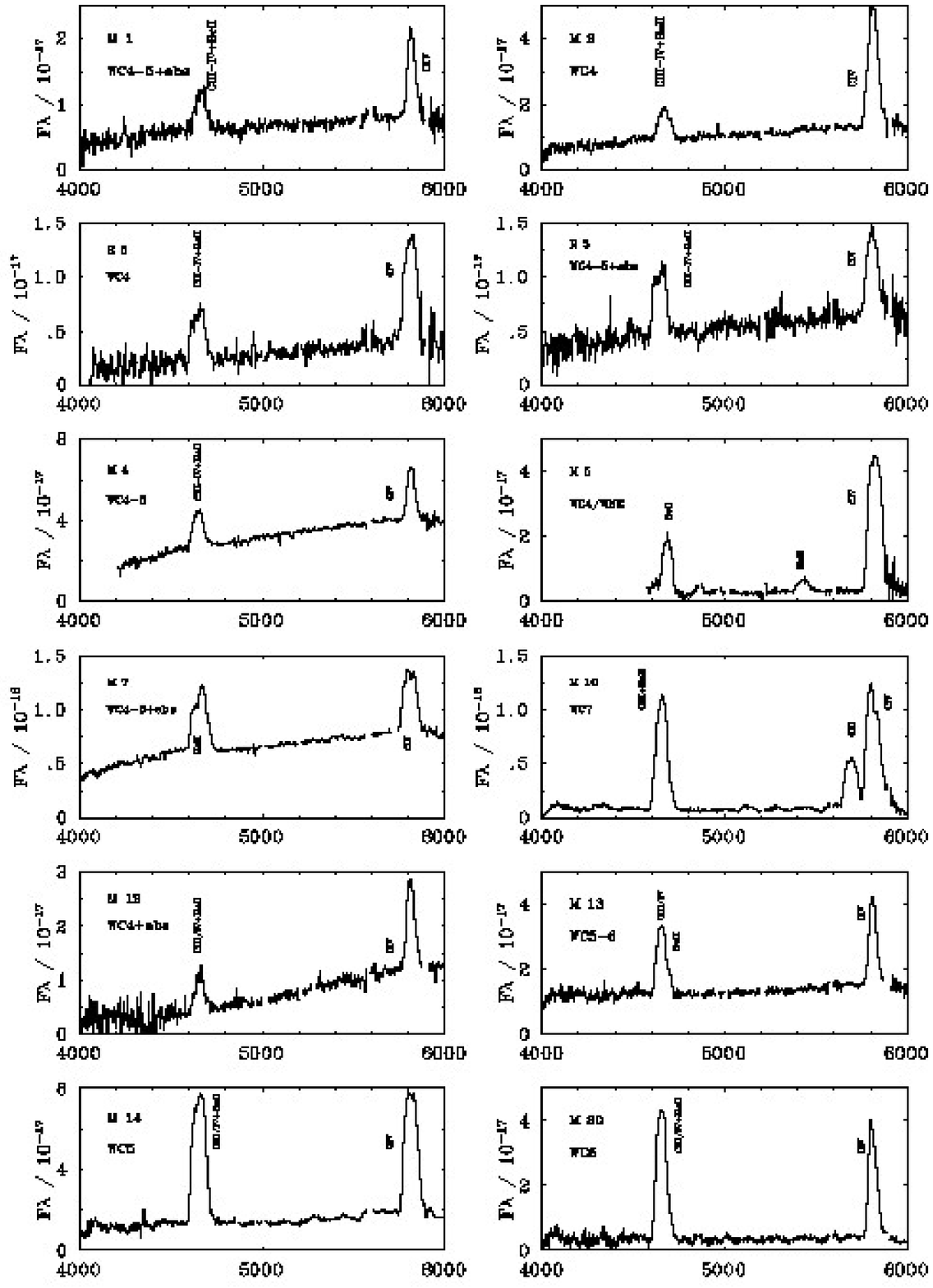}
\caption{Flux calibrated (erg\,cm$^{-2}$\,s$^{-1}$\,\AA$^{-1}$) optical 
spectroscopy of WN/C and WC stars in IC~10 
observed with Gemini GMOS.}
\label{WC}
\end{figure*}

Absolute magnitudes, $M_{v}$, shown in Table~\ref{wrcat}
then follow from an adopted distance modulus (DM) of 23.86 ($d$=590 kpc),
as discussed below. At this distance one arcsec subtends $\sim$3\,pc.
Interstellar reddenings and absolute magnitudes  are rather uncertain,  
but typical values compare well with other Local Group WR stars (e.g. van 
der Hucht 2001). In general, stars which appear to be multiple 
spectroscopically, are  indeed intrinsically bright, whilst those which 
may be single are faint, with the exception of RSMV 8 for which 
$M_{v}$(WN10)=$-$7.6 mag is realistic when compared with other very late 
WN stars (Crowther \& Smith 1997).

Our adopted distance was derived from near-IR photometry by Borissova et 
al. (2000), with respect to IC~1613, which possesses a more robust 
distance modulus (Freedman et al. 2001). This lies between the
two distance moduli obtained by Sakai et al. (1999) from
Cepheid variables (DM=24.1, $d$=660 kpc) and from the tip of the RGB, 
(DM=23.5, $d$=500 kpc). Clearly, the high and variable extinction towards
IC~10 favours IR techniques which are less susceptible to reddening.
Our present, imprecise, method of determining reddenings favours the
Borissova et al. distance, whilst lower reddenings (e.g. from nebular 
methods) may favour higher distance scales (e.g. DM=24.57, $d$=820 kpc: 
Kennicutt et al. 1998).




\subsection{WN stars}

We first discuss two stars that have previously been 
classified as WN stars, yet their exact nature is in doubt, 
given their weak reported He\,{\sc ii} emission line strengths.
Ultimately, the dividing line between O3\,If/WN stars, 
WN9--11 stars and normal WN stars is rather arbitrary, with 
$W_{\lambda}$(He\,{\sc ii} 4686)$>$10--12\AA\ in most cases (see 
Crowther \& Dessart 1998). Nevertheless, one needs to ensure consistency 
when comparing the population of emission line stars in external 
galaxies, or else e.g. WC/WN ratios would be artificially skewed 
in favour of WN stars.  

[MAC92] 17 is classified as a WN star according to Massey \& Armandroff
(1995), despite `very weak emission' at He\,{\sc ii} 4686.
Our initial GMOS extraction confirmed He\,{\sc ii} 
$\lambda$4686 emission, albeit with  an equivalent width of 
only $\sim$8\AA, such that [MAC92] 17 most closely resembled an
O3\,If/WN star. Alternatively, the stellar continuum
of a genuine WN star may be heavily contaminated by a line-of-sight 
or binary companion. One might suspect the latter case given the large 
distance to IC~10, so we have inspected HST/WFPC2 images which reveal 
two sources, -A and -B,  within 1$''$ of the coordinates of 
Massey \& Holmes (2002, see Fig.~\ref{chart2}). 
Therefore, we subsequently
re-examined our raw frames and discovered that He\,{\sc ii} emission 
spanned only half of the stellar continuum.  We carefully re-extracted this
narrow ($\sim$5 pixel) region, which now revealed a significantly stronger He\,{\sc ii} $\lambda$4686
emission of $\sim$30\AA, with weak N\,{\sc v} $\lambda\lambda$4603--30 emission
also present. This spectrum in the vicinity of $\lambda$4686 is presented in Fig.~\ref{ofwn}
and so [MAC92] 17-A is clearly consistent with a WNE star.


There is  another weak He\,{\sc ii} emission line star in IC~10, which
has recently been classified as a WN star, namely WR24 from Massey \& Holmes
(2002), hereafter [MAC92] 24  (following the CDS nomenclature -- see 
Sect~3.1). 
We have not obtained new spectroscopy of this star in the present 
program, given its location within the bright H\,{\sc ii} region HL111c (Hodge \& Lee 
1990)\footnote{
Note that the putative WN star identified by
Richer et al. (2001) in HL111c is most likely [MAC92] 24. Richer et al. described a WN star with
$W_{\lambda}$(He\,{\sc ii} 4686)=3\AA, consistent with Massey \& Holmes (2002),
and their coordinates agree to within {\it 2 arcsec}, comparable to the slit width used
in the former.}. [MAC92] 24 was not identified as a He\,{\sc ii}
emission line source in Royer et al. (2001) and its nature is also in doubt.

Spectroscopically, the published data for [MAC92] 24 most closely
resembles an O3\,If\,WN star, such as Melnick 42 in the LMC, which
has a higher He\,{\sc ii} $\lambda$4686 equivalent width, plus a similar FWHM.
Massey \& Holmes (2002)
favoured a WN binary nature given that it was exceptionally bright.
Therefore, we have again inspected HST/WFPC2 F555W images, which reveals 
three sources within 1$''$  from the published coordinates of [MAC92] 24, as indicated in
Fig~\ref{chart2}. The ground-based slit spectrum will have contributions
from each, with the continuum dominated by source A (V$\sim$18.8 mag).
All are consistent
with early-type stars, with similar U--V colours, preventing any further 
discrimination of the true emission line source. The severe crowding of this region is
apparent from the figure, with two further bright sources lying 
within 2$''$, to the N (V$\sim$19.4 mag) and NE (V$\sim$19.8 mag). 
From a combination of the existing spectroscopy and imaging, one would favour a WN+OB binary, or a 
perhaps a cluster containing a WN star. 

If this is the case, one puzzle remains. The high S/N spectroscopy 
of [MAC92] 24 presented by Massey \& Holmes (2002) shows no sign of N\,{\sc iii-v} 
$\lambda\lambda$4603--40 emission, which would be expected if this was 
indeed a WN+OB binary. Undoubtedly, higher spatial resolution imaging/spectroscopy 
is needed for WR confirmation of this object, but for the moment we adopt 
WN+OB for its spectral type despite the lack of N\,{\sc iii-v} emission.



Let us now turn to the more conventional WN stars in IC~10.
Figure~\ref{WN} presents Gemini GMOS spectroscopy of eight WN stars.
Precise WN spectral types are rather difficult to establish since 
spectroscopy for most stars is limited to regions redward of N\,{\sc iv} 
$\lambda$4058. Nevertheless, including  [MAC92] 23 from Massey \& Holmes (2002), 
IC~10 hosts seven early (2--5) WN stars and three late (6--11) WN stars, 
plus [MAC92] 24 for which we have no subtype information. 
Amongst the WNE stars, He\,{\sc ii} $\lambda$4686  emission is extremely weak in RSMV 9 and 
RSMV 12.  The latter might be a binary. WFPC2 imaging reveals no 
nearby line-of-sight companions contaminating our Gemini spectroscopy. 
The  remainder are  potentially single, and have line strengths and widths comparable to 
LMC WNE stars. RSMV 2 and [MAC92] 23\footnote{Royer et al. (2001) 
identified [MAC92] 23 as a strong 
He\,{\sc ii} $\lambda$4686 emission source, but it was excluded from their catalogue on the 
basis of a very red colour (c1--c2$>$1 mag), common to many foreground late-type stars.}
are late WN7--8 stars with relatively strong N\,{\sc iii} $\lambda\lambda$4634--40
emission, although the former has a faint companion as revealed by WFPC2 imaging.
RSMV 8 is another unusual example of a very late WN10 star  (e.g. Crowther \& Smith 1997).

For comparison, the Solar Neighbourhood hosts a similar fraction of 
WNE and WNL stars (van der Hucht 2001), whilst WNE stars outnumber WNL stars by
approximately 5:2 in the LMC  (Breysacher et al. 1999),
and there are no WNL stars in the SMC (Massey \& Duffy 2001), such that
the WN population of IC~10 is closer to the LMC than the SMC. Crowther (2000)
explained the general trend to earlier spectral type amongst WN stars at
lower metallicity due to the differing dependence of nitrogen emission lines
on reduced CNO equilibrium abundances.  Indeed, in contrast with Galactic
late WN stars, there are no IC10 cases in which N\,{\sc iii} $\lambda\lambda$4640 
is comparable in  strength to He\,{\sc ii} $\lambda$4686, an effect remarked upon already 
for the LMC by Crowther \& Smith (1997).

\subsection{WN/C and WC stars}

\begin{figure}[hbtp]
\epsfxsize=8.8cm \epsfbox[25 510 280 735]{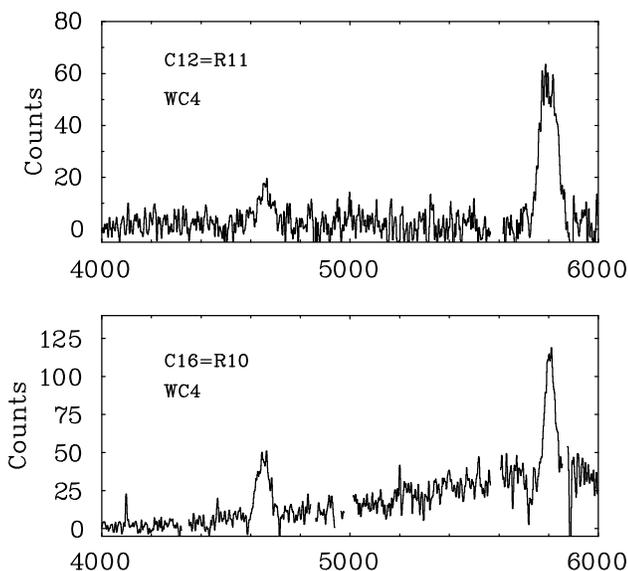}
\caption{Optical spectroscopy of WC stars in IC~10 
observed with CFHT MOS.}
\label{WC2}
\end{figure}

Figure~\ref{WC} presents Gemini GMOS spectrophotometry of twelve WC stars, including 
[MAC92] 5 which is newly identified here as a WNE/WC4 star, whilst 
Fig.~\ref{WC2} shows MOS spectroscopy for RSMV 10 and 11.
It is well known that
late type (WC8--9) stars are only observed in metal-rich environments, such
as the inner Milky Way and M31 (van der Hucht 2001). Indeed, of the 14 WC
stars in IC~10, at least twelve are WC4--6 subtypes. [MAC92] 10
is a notable exception, i.e. a WC7 star with extremely broad lines
($v_{\infty}\sim$  3000 km\,s$^{-1}$). 
[MAC92] 15 was listed as WC6--7 by Massey \& Armandroff (1995), though
their Figure~1 appears to rule out a subtype as late as WC7 for this star. 
Consequently, [MAC92] 10 appears to be {\it unique}  amongst low metallicity WC stars in having 
relatively strong
C\,{\sc iii} $\lambda$5696 emission. Crowther et al. (2002) recently 
claimed to have 
established the origin of the WC subtype distribution amongst Local
Group galaxies. They established a metallicity dependence amongst WC winds,
reminiscent of OB stars, with the strength of C\,{\sc iii} $\lambda$5696 
exceptionally sensitive to wind density. The dominance of WCE stars within
the low metallicity environment of IC~10 naturally follows this pattern,
with the conspicuous exception of [MAC92] 10, which is remarkably similar to
V378 Vul (alias WR125) in our Galaxy. 

\begin{figure} 
\vspace{7.5cm}
\includegraphics{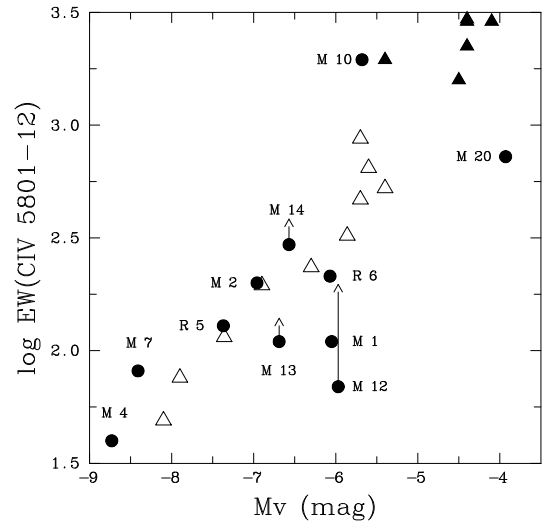}
\caption{Comparison between the emission C\,{\sc iv} $\lambda$5801-12 emission
equivalent widths (in \AA) measured in our Gemini and CFHT datasets 
and absolute visual magnitude estimates for IC~10 WC stars (filled circles). 
Corrections to measured GMOS equivalent widths are shown for three stars, owing to
contamination by nearby companions from HST WFPC2 images.
Single (filled triangles) and binary (open triangles) WC stars
in the LMC are shown for comparison (taken from Smith et al. 1990) 
assuming an LMC distance modulus of 18.50.}
\label{WCplot}
\end{figure}

In  contrast with the WN population, the majority of WC stars are 
heavily contaminated by OB line-of-sight or binary light. 
This is demonstrated in Fig.~\ref{WCplot} where we compare the
C\,{\sc iv} $\lambda$5801--12 emission line strengths and estimated
absolute visual magnitudes from our ground-based photometry of IC~10
WC stars versus LMC WC single star and binaries (Smith et al. 1990).
From this comparison, [MAC92] 10 and 20 appear to be convincingly single, 
or at least the lines and continua are not strongly diluted. All others
suffer considerable contamination in their spectra. More IC~10 WC 
stars may be single, but higher angular resolution observations would
be required. Indeed, corrections for line-of-sight close companions 
are shown in Fig.~\ref{WCplot} for [MAC92] 12, 13 and 14 based on WFPC2
imaging.

Nevertheless, cases such as [MAC92] 10 and 20 are important, since
it is commonly assumed that binary evolution (via Roche Lobe overflow)
is {\it required} for evolution to late Wolf-Rayet phases at 
low metallicity. For instance, the only carbon/oxygen sequence WR 
star in the SMC, Sand 1 (WO+O4V), has a
close massive companion (Moffat et al. 1985). Therefore, the presence
of (at least) two such cases in IC~10 suggests that progression 
to late phases via {\it single} stellar evolution may be achieved at 
low metallicities.  The difference with respect to the SMC may solely 
be as a result of a much larger massive stellar population, such that there
will be a higher statistical likelyhood of extremely 
massive stars in IC~10, under the assumption of a universal Salpeter IMF.

Finally, we have compared the C\,{\sc iv} $\lambda\lambda$5801--12 line
flux in IC~10 WC stars with those of LMC WC stars. Smith et al. (1990)
claimed that LMC early type WC stars possessed uniform 
$\lambda\lambda$5801--12 line fluxes of 
$\log F_{\lambda} = -7.6 \pm 0.1$ 
erg\,s$^{-1}$\,cm$^{-2}$. Despite the imprecise reddenings to many IC~10
WC stars, we obtain an essentially identical mean line flux, 
$\log F_{\lambda} = -7.5^{+0.2}_{-0.6}$ erg\,s$^{-1}$\,cm$^{-2}$. Firmer
conclusions relating to a universal early WC await more reliable reddening
determinations from HST.




\begin{figure} 
\vspace{7.5cm}
\includegraphics{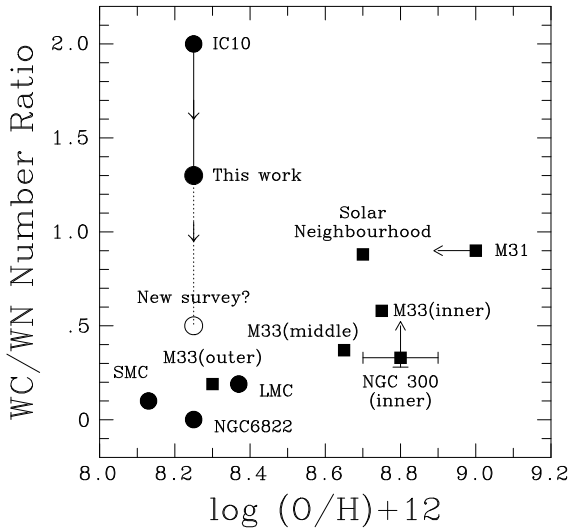}
\caption{WC/WN ratio for Local Group and Sculptor Group 
spiral (squares) and irregular  galaxies (circles) versus oxygen content.
Individual data points are taken from Massey \& Johnson (1998), Schild et al.
(2003).  Recent evidence,  based on model atmosphere calculations 
rather than H\,{\sc ii} region analyses, suggests a lower metallicity for M31 (Trundle 
et al. 2002). The ratio for IC~10 is shown  based on the Massey \& Johnson (1998) 
and present spectroscopic results (filled circles), plus a 
re-assessment of recent narrow-band imaging of Massey \& Holmes  (2002, open circle).}
\label{WC_WN}
\end{figure}

\section{WC/WN ratio for IC~10}\label{ratio}

What effect does our new spectroscopic results have on the unusual 
WC/WN ratio for IC~10?
Our datasets, together with previous spectroscopy
imply WC/WN=14/11$\sim$1.3, where we have omitted the
intermediate star [MAC92] 5 from this ratio. Comparison with the WC/WN ratios for other
Local Group galaxies in Fig.~\ref{WC_WN} would lead one to conclude that 
WC/WN$\sim$0.15 is anticipated for IC~10.
Clearly, for IC~10 to mimic other Local Group galaxies, another $\sim$80
WN stars would be expected. According to WHT imaging by 
Royer et al. (2001) this is not likely, {\it but} 
Massey \& Holmes (2002) claim up to $\sim$75  WR stars still await 
confirmation. 
  
One has to bear in mind that the exact number expected depends heavily upon
the strength of the excess measured in the WR filter relative to the continuum
filter by Massey \& Holmes (2002). For example, examination of their
Fig.~3 implies that $\sim$15 have excesses greater than 0.5 mag, most of which
{\it are} probably genuine WN stars, from comparison with their control field. 
Taking 
these into account would reduce the WC/WN ratio to $\sim$0.5, as indicated
with an open circle in Fig.~\ref{WC_WN}. This would make IC~10 unusually rich in
WC stars for a low metallicity galaxy. In contrast, Massey \&  Holmes
were rather bolder, claiming instead WC/WN$\sim$0.3. This was on the basis
that {\it all} stars with excesses between 0.3--0.5 mag are bona fide WR 
stars, even though several stars in their control field had such excesses.
Their justification for this conclusion was heavily reliant on [MAC92] 24 
being a genuine WN star, since its excess lay at the  $\sim$0.3 mag limit.
As we have discussed above, this object would not ordinarily
be classified as a genuine WR star from a spectroscopic viewpoint, since (individual)
stars with an emission equivalent width of $\sim$4\AA\ at  He\,{\sc ii} $\lambda$4686 
are most likely to be Of supergiants. In the case of [MCA92] 24, we consider it to be 
a probable WN star on the basis of additional information afforded by HST imaging, 
namely its brightness and apparent multiplicity. Consequently,  a sizeable fraction of 
stars in IC~10 with such a  small excess identified by Massey \& Holmes are likely to be  
Of,  Of/WN stars or spurious candidates. 

In fairness, the situation discussed above for IC~10 is common to all 
galaxies beyond the Magellanic Clouds, such that one has to be wary when comparing WC/WN ratios.
There are cases of spectroscopically confirmed WRs stars in other Local Group galaxies
with magnitude differences as low as 0.2 mag, although the
vast majority greatly exceed 0.5 mag (e.g. Fig.~4 of Massey  \& Johnson 1998).
A handful of genuine WR stars {\it are} known for which He\,{\sc ii} emission strengths do not exceed
10--12\AA, namely: (i) WR+OB binaries with extreme light ratios, such as Sk\,108 in the SMC
(e.g. Foellmi et al. 2003), although it should be noted that this is the {\it only} 
WN+OB binary in the SMC for which He\,{\sc ii} $\lambda$4686 does not exceed 10\AA; (ii)
very late WN stars, such as RSMV 8 in IC~10, have very small He\,{\sc ii} $\lambda$4686 emission 
equivalent widths, which are similar to extreme Of stars (Crowther \&  Bohannan 1997).  
Such stars are, generally, rather rare relative to Of stars. 

For the moment, the jury remains out until all new candidates from
Massey \& Holmes (2002) are spectroscopically  observed. 
Ideally, the presence of photospheric absorption lines
would be sought to resolve current uncertainties. More realistically, given  the visual 
faintness and slit contamination, the strength of emission line fluxes might be 
used to verify WR versus Of/WN  identifications.  For the moment, we merely suggest that 
the WC/WN ratio of  IC~10 may not be as low as $\sim$0.3, as has been claimed by Massey 
\& Holmes (2002). Consequently, it is possible that the  number and distribution of the 
WR population  for IC~10 remains unusually high, relative to other Local Group irregular 
galaxies, as further evidence of a short,  co-eval burst of star formation in the recent 
past. 

In summary, we show that IC~10 hosts a substantial population of Wolf-Rayet
stars. At present, spectroscopically confirmed WC stars outnumber WN 
stars, and early subtypes of both flavours outnumber late subtypes. 
Recent candidate WC9 stars are not  confirmed, and  remain exclusive to 
metal-rich environments, although the  presence of
[MAC92] 10, an apparently single, broad lined WC7 star was not anticipated in
an environment of 0.25$Z_{\odot}$. This star illustrates that highly
evolved Wolf-Rayet populations can occur in low metallicity environments
without mass transfer in a close binary. WN stars are reminiscent of the LMC,
with a large WN3--4 population, plus a few late ([MAC92] 23, WN7--8) and very 
late (RSMV 8, WN10) subtypes. Work is underway to analyse a subset of these objects,
for comparison with counterparts in Local Group galaxies spanning a range
of metallicities. Spectroscopy of all potential candidates from
Massey \& Holmes (2002) is keenly sought in order to resolve remaining questions.


\begin{acknowledgements} 
Based on observations obtained at the Gemini Observatory, which is operated by the Association of Universities for Research in Astronomy, Inc., under a cooperative agreement with the
       NSF on behalf of the Gemini partnership: the National Science Foundation (United States), the Particle Physics and Astronomy Research Council (United Kingdom), the National
       Research Council (Canada), CONICYT (Chile), the 
Australian Research Council (Australia), CNPq (Brazil), and 
CONICET (Argentina).  Part of the data were obtained with the 
Canada-France-Hawaii Telescope,
  which is operated by the National Research Council of Canada,
  the Centre National de la Recherche Scientifique de France, and the
  University of Hawaii.
Thanks to 
Inger Jorgensen for help with extraction of the Gemini datasets,
to Pierre Martin for his contribution to the CFHT observations, 
and to Cedric Foellmi for providing spectroscopy of Sk\,108 and Melnick 42.
PAC and SJS acknowledge financial support from the Royal Society
and PPARC respectively, and LD acknowledges NSERC and the
Canada Research Chair Program
\end{acknowledgements}

\end{document}